\documentclass[prd,aps, tightenlines, preprintnumbers, showpacs, nofootinbib,superscriptaddress,notitlepage,11pt]{revtex4-1}

\usepackage{epsfig}
\usepackage{amsmath,amsthm,amssymb}
\usepackage{mathtools}
\usepackage{siunitx}

\usepackage[usenames,dvipsnames]{color}

\newcommand{\Slash}[1]{{\ooalign{\hfil/\hfil\crcr$#1$}}}

\newcommand{\beq}{\begin{eqnarray}}
\newcommand{\eeq}{\end{eqnarray}}

\newif\iffigure
\figuretrue

\begin{document}
\preprint{YITP-18-67}

\title{Probing the Weizs$\ddot{{\bf a}}$cker-Williams gluon Wigner distribution in $pp$ collisions }

\author{Renaud Boussarie}
\affiliation{Institute of Nuclear Physics, Polish Academy of Sciences, Radzikowskiego 152, PL-31-342 Krak$\acute{o}$w, Poland}

\author{Yoshitaka Hatta}
\affiliation{Yukawa Institute for Theoretical Physics, Kyoto University, Kyoto 606-8502, Japan}

\author{Bo-Wen Xiao}
\affiliation{Key Laboratory of Quark and Lepton Physics (MOE) and Institute
of Particle Physics, Central China Normal University, Wuhan 430079, China}
\affiliation{Centre de Physique Th\'eorique, \'Ecole Polytechnique, 
CNRS, Universit\'e Paris-Saclay, Route de Saclay, 91128 Palaiseau, France.}

\author{Feng Yuan}
\affiliation{Nuclear Science Division, Lawrence Berkeley National
Laboratory, Berkeley, CA 94720, USA}

\begin{abstract}
We show that the diffractive forward production of two quarkonia, especially the $\chi_{cJ}$ and $\eta_c$ states, in proton-proton or proton-ion collisions can access the Weizs$\ddot{{\rm a}}$cker-Williams gluon Wigner distribution of the proton. We use the  hybrid factorization  approach where the collinear, double gluon PDF is applied for one of the protons and the $k_T$-dependent (Wigner or GTMD) distribution for the other. The production of quarkonia is treated in the NRQCD framework. A particularly concise formula is obtained for double $\chi_{J=1}$ production.  
\end{abstract}
\maketitle

\section{Introduction}

In hadron physics research, one of the ultimate goals is to depict the colorful and kaleidoscopic multi-dimensional landscape of the internal structure of hadrons including 
nucleons and nucleus. In particular, in addition to the longitudinal momentum distribution of partons inside hadrons as given by the Feynman parton distributions, 
we also intend to learn about the transverse spatial (generalized parton distributions (GPD)) and transverse momentum (transverse momentum dependent (TMD)) distributions. The so-called quantum phase space Wigner distributions \cite{Ji:2003ak, Belitsky:2003nz,Lorce:2011kd}  encode all the above important informations and are viewed as the mother distributions of all. In practice, since it is rather difficult to directly measure the spatial transverse coordinates in high energy scatterings, we also normally define the generalized transverse momentum distribution (GTMD)\cite{Meissner:2009ww, Lorce:2013pza, Echevarria:2016mrc} as the Fourier transform of the corresponding Wigner distribution for quarks and gluons. 

Interestingly, as pointed out in Refs.~\cite{Mulders:2000sh,Bomhof:2006dp, Xiao:2010sp, Dominguez:2010xd, Dominguez:2011wm}, quark and gluon TMDs are not unique 
due to different possible structures of gauge links representing the initial and final state interactions. As far as the gluon TMDs are concerned, there are at least two nontrivial gluon distributions whose difference becomes crucial especially in the small-$x$ region. One is known as the dipole gluon distribution which contains the dipole-shaped configuration of gauge links, while the other is called the Weizs$\ddot{{\rm a}}$cker-Williams (WW) gluon distribution which features either a future-pointing or past-pointing gauge link, but not both.
Naturally, one expects that this kind of complication should also persist in the case of Wigner distributions and GTMDs. 

Generally speaking, up to now, the HERA experiment has provided us with the most precise knowledge of Feynman parton distributions for a large range of $x$ region, through the inclusive deep inelastic scattering (DIS) process. Furthermore, less inclusive processes, such as the semi-inclusive DIS (SIDIS) and the deeply virtual Compton scattering (DVCS) can help us gain insights on the TMDs and GPDs, respectively. The important question here is whether one can probe the Wigner distributions or GTMDs experimentally. 
It is rather challenging to achieve such a goal because, obviously, one has to consider  more exclusive/complicated processes than SIDIS and DVCS. 

Recently, we have found that the goal of measuring the gluon Wigner distributions can be achieved at the future electron-hadron colliders \cite{Hatta:2016dxp}, such as the planned electron-ion colliders (EIC) and the large hadron electron collider (LHeC) \cite{Boer:2011fh, AbelleiraFernandez:2012cc, Accardi:2012qut}. 
Specifically, the {\it dipole} gluon Wigner or GTMD can be probed in diffractive dijet process in the small-$x$ region at EIC \cite{Aschenauer:2017jsk}.
 Subsequently, there have been a lot of progress on this topic in the last two years \cite{Boussarie:2016ogo, Hagiwara:2016kam, Zhou:2016rnt, Ji:2016jgn, Hatta:2016aoc, Hagiwara:2017ofm, Bhattacharya:2017bvs, Hatta:2017cte, Hagiwara:2017fye, More:2017zqp, {Raja:2017xlo},Hagiwara:2017uaz}. Especially,  one can access and study the extremely evasive gluon orbital angular momentum with the help of the spin-dependent gluon GTMD as discussed in Refs.~\cite{Hatta:2011ku,Zhao:2015kca, Ji:2016jgn, Hatta:2016aoc,Bhattacharya:2017bvs,Bhattacharya:2018lgm,Raja:2017xlo}. 

On the other hand, it has been realized already in Ref.~\cite{Hatta:2016dxp} that the Weizs$\ddot{{\bf a}}$cker-Williams Wigner (WWW) distribution is more difficult to access. One possibility, as suggested in \cite{Hatta:2016dxp}, was to look at a process involving two incoming photons and four outgoing jets at EIC. However, the transition rate would be too small to make it a realistic measurement. The objective of this paper is to show that it is possible to rather directly probe the WWW distribution and WW type GTMD for small-$x$ gluons in the production of two heavy scalar quarkonia such as $\chi_c$ and $\eta_c$ in the forward rapidity region in diffractive $pp$ and $pA$ collisions $pp, pA \to \chi_c \chi_c pX$.

The rest of the paper is organized as follows. We first introduce the WW gluon GTMD and briefly describe the process we will consider in this paper, i.e., double quarkonia production in the hard diffractive processes in $pp$ and $pA$ collisions. In Sec.~III, we derived the amplitude for the double quarkonia production through double gluon scattering on the nucleon/nucleus targets. These amplitudes will be converted into $pp$/$pA$ collisions cross sections in Sec. IV by applying the double parton scattering framework for the incoming two gluons from the projectile. In Sec.III and IV, we derive the results for both $\eta$ and $\chi$. In particular, we will present the explicit differential cross sections on different combinations of two quarkonia states of $\eta$, $\chi_{0,1,2}$. Finally, we summarize our paper in Sec. VI.\\



\section{Weizs$\ddot{{\bf A}}$cker-Williams gluon GTMD}

We start by writing down the Weizs$\ddot{{\rm a}}$cker-Williams (WW) gluon GTMD. For the `target' proton fast-moving in the negative $z$-direction we define
\begin{equation}
x\mathcal{G}^{ij}\left(\boldsymbol{K},\boldsymbol{\Delta}\right)\equiv 2\int\frac{d^{3}z}{P^-\left(2\pi\right)^{3}}e^{ixP^-z^+  -i\boldsymbol{K}\cdot\boldsymbol{z}}\left\langle P-\frac{\Delta}{2}\left|\mathrm{Tr}\left[U^\dagger_\pm F^{-i}(z/2) U_\pm F^{-j}(-z/2) \right]\right|P+\frac{\Delta}{2}\right\rangle, \label{eq:GTMDdef}
\end{equation}
where $U_\pm$ is the staple-shaped fundamental Wilson line connecting the points $z/2$ and $-z/2$ via light-like Wilson lines encircling $x^+=\pm \infty$. 
 Boldface letters denote two-dimensional  vectors. We have kept the transverse indices $i,j=1,2$ open because this is what we shall need in later calculations. In the small-$x$ region, we may approximate $e^{ixP^-z^+} \approx 1$ and find (c.f., \cite{Dominguez:2011wm})
\begin{equation}
x\mathcal{G}^{ij}\left(\boldsymbol{K},\boldsymbol{\Delta}\right)\approx - \frac{2}{\alpha_s} \int\frac{d^{2}\boldsymbol{b}_{1}d^{2}\boldsymbol{b}_{2}}{\left(2\pi\right)^{4}}e^{-i\boldsymbol{\Delta}\cdot\frac{\boldsymbol{b}_{1}+\boldsymbol{b}_{2}}{2}-i\boldsymbol{K}\cdot\left(\boldsymbol{b}_{1}-\boldsymbol{b}_{2}\right)}\frac{\left\langle P-\frac{\Delta}{2}\left|\mathrm{Tr}\left[\left(\partial^{i}U_{\boldsymbol{b}_{1}}^{\dagger}\right)U_{\boldsymbol{b}_{1}}\left(\partial^{j}U_{\boldsymbol{b}_{2}}^{\dagger}\right)U_{\boldsymbol{b}_{2}}\right]\right|P+\frac{\Delta}{2}\right\rangle }{\left\langle P|P\right\rangle },\label{eq:GTMDdef}
\end{equation}
where $U_{\boldsymbol{b}}$ is the lightlike Wilson line from $x^+ = -\infty$ to $x^+=+\infty$ at fixed transverse position $\boldsymbol{b}$.
It is easy to check from (\ref{eq:GTMDdef}) the relations $\mathcal{G}^{ij}\left(\boldsymbol{K},\boldsymbol{\Delta}\right)=\mathcal{G}^{ji}\left(-\boldsymbol{K},\boldsymbol{\Delta}\right) = (\mathcal{G}^{ij}(-\boldsymbol{K},-\boldsymbol{\Delta}))^*$. The general parameterization is thus\footnote{Naively, there is another term proportional to the tensor structure  $\boldsymbol{K}^i\boldsymbol{\Delta}^j + \boldsymbol{K}^j\boldsymbol{\Delta}^i-\boldsymbol{K}\cdot \boldsymbol{\Delta}\delta^{ij}$. However, as pointed out in \cite{Boer:2018vdi}, this term is not independent and can be absorbed in ${\cal G}_{2,3}$. See, also, \cite{Lorce:2013pza}. } 
\begin{equation}
\mathcal{G}^{ij}\left(\boldsymbol{K},\boldsymbol{\Delta}\right)\equiv\delta^{ij}\mathcal{G}_{1}+\left(\frac{\boldsymbol{K}^{i}\boldsymbol{K}^{j}}{\boldsymbol{K}^{2}}-\frac{\delta^{ij}}{2}\right)\frac{\boldsymbol{K}^{2}}{M^{2}}\mathcal{G}_{2}+\left(\frac{\boldsymbol{\Delta}^{i}\boldsymbol{\Delta}^{j}}{\boldsymbol{\Delta}^{2}}-\frac{\delta^{ij}}{2}\right)\frac{\boldsymbol{\Delta}^{2}}{M^{2}}\mathcal{G}_{3}+\left(\frac{\boldsymbol{K}^{i}\boldsymbol{\Delta}^{j}-\boldsymbol{\Delta}^{i}\boldsymbol{K}^{j}}{M^{2}}\right)\mathcal{G}_{4},\label{eq:GTMDdef-1}
\end{equation}
where $M$ is the nucleon mass. ${\cal G}_{1,2,3,4}$ are all real and depend on $\boldsymbol{K}^2$, $\boldsymbol{\Delta}^2$ and $(\boldsymbol{K}\cdot \boldsymbol{\Delta})^2$. 
In the forward limit $\boldsymbol{\Delta}=0$, ${\cal G}_1$ and ${\cal G}_2$ reduce to the unpolarized and linearly polarized WW gluon TMDs, respectively. 
The $\boldsymbol{K}$-integral of ${\cal G}_3$ is proportional to the gluon transversity GPD which in turn is related to the so-called elliptic gluon Wigner distribution \cite{Hatta:2017cte}.    From what we know about the latter distribution \cite{Hatta:2016dxp,Hagiwara:2016kam,Hagiwara:2017ofm}, we presume that ${\cal G}_3$ is numerically small, on the order of a few percent  effect compared to ${\cal G}_{1,2}$.  

\begin{figure}
\begin{centering}
\includegraphics[scale=0.6]{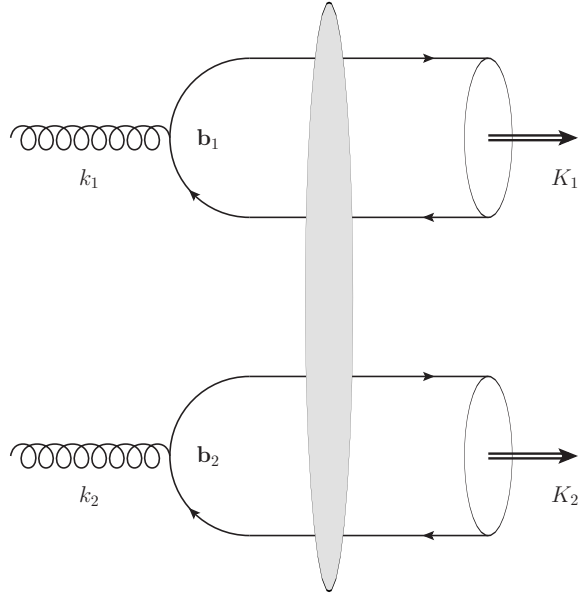}
\par\end{centering}
\caption{Production of 2 quarkonia from 2 gluons \label{fig1}}
\end{figure}

Various experimental processes have been identified to probe the WW gluon TMD \cite{Dominguez:2011wm}. On the other hand, the measurement of the GTMDs is more challenging, and most of the proposals so far concern the dipole gluon GTMD whose treatment is somewhat simpler because of its relation to the dipole S-matrix   \cite{Hatta:2016dxp}. In this paper, we show that  double quarkonium production in $pp$ collisions $pp \to M_1M_2 pX$ is a very sensitive observable for  the WW gluon GTMD. This is a natural extension of the previous observation  \cite{Qiu:2013qka} (see, also, \cite{Kang:2013hta}) that the WW gluon TMD can be probed in single quarkonium production $pp\to MX$. 
The hard subprocess and the full process are respectively depicted in Fig.~\ref{fig1} and in Fig.~\ref{fig3}. Two gluons are emitted from the projectile proton and scatter off the shockwave field created by the target proton. In the final state, we measure two quarkonia with  momentum $K_1$ and $K_2$ in the forward region as well as the elastically scattered target with momentum transfer $\boldsymbol{\Delta}=\boldsymbol{K}_1+\boldsymbol{K}_2$ in the backward region. To lowest order, the shockwave consists of two gluons in the $t$-channel, and we can view this diagram as the square of  the $gg^*\to M$ amplitude. This means that the produced quarkonia must be $C$-even states, such as $\eta_{c}$ and $\chi_{cJ}$ (and their bottomonium counterparts). We study this process in the `hybrid factorization' approach (collinear gluons from one proton and non-collinear gluons from the other proton) and show that the cross section  can be written as a convolution of the square of the WW gluon GTMD of the target and the double gluon PDF of the projectile.

  It should be mentioned that a very similar idea came out  recently in Ref.~\cite{Bhattacharya:2018lgm} where the authors proposed to measure the gluon GTMD (including its spin dependence) in doubly-diffractive double $\eta_c$ production $pp\to pp\eta_c\eta_c$ for moderate values of $x$. Their argument is limited to the two-gluon exchange level where the difference between WW and dipole distributions becomes immaterial. Our process is single-diffractive and we do not consider spin effects. Instead,  we focus on the forward, small-$x$ region and carefully examine the structure of the relevant Wilson lines. Our result suggests that the GTMD discussed in  \cite{Bhattacharya:2018lgm} likely becomes the WW GTMD once higher order rescattering effects are taken into account. 
  

\section{Quarkonium production amplitude}

\subsection{Single quarkonium production}

\begin{figure}
\begin{centering}
\includegraphics[scale=0.75]{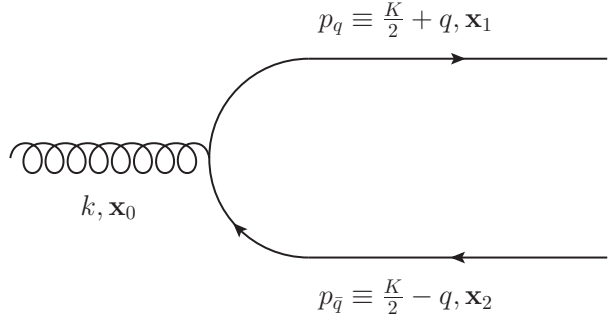}
\par\end{centering}
\caption{Open charm production \label{fig2}}
\end{figure}

\begin{figure}
\begin{centering}
\includegraphics[scale=0.5]{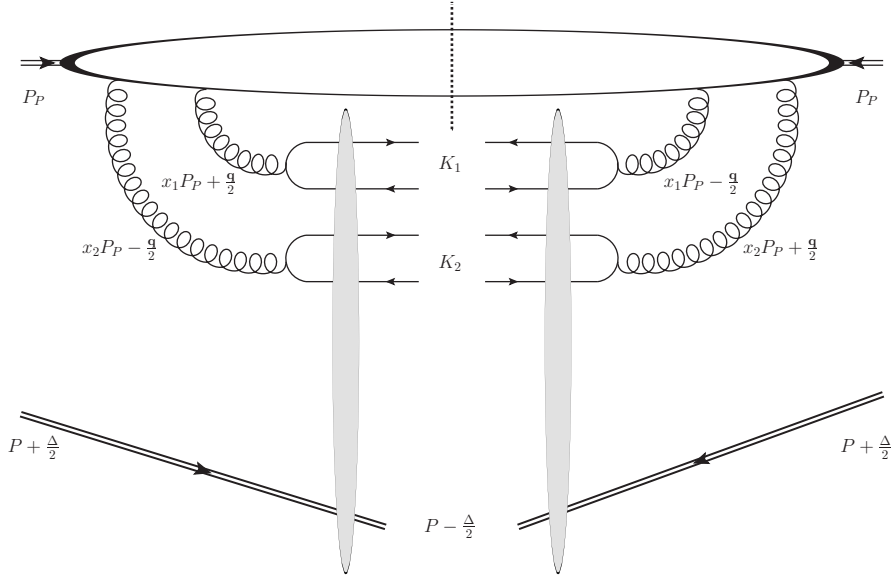}
\par\end{centering}
\caption{Production of a quarkonium pair in hybrid factorization with double scattering on the projectile side}\label{fig3}
\end{figure}

As a warm-up, we first compute the amplitude for the single quarkonium production in the hybrid factorization approach combined with the non-relativistic QCD (NRQCD) framework.  Related calculations can be found in the literature, e.g., \cite{Qiu:2013qka,Kang:2013hta}.   
The right-moving projectile proton is treated as a dilute object, i.e. partons
are extracted from the projectile using regular Parton Distribution
Functions, whereas the left-moving target is treated as a showckwave. 
The subamplitude for the open production of a massive $q\bar{q}$
pair from a gluon (see Fig.~\ref{fig2}) can be straightfowardly written as, in $D=d+2$ dimensions,\footnote{ 
Our notation is as follows: 
We define two lightlike vectors $n_{1},n_{2}$ such that the projectile
flies along $n_{1}$ and the target flies along $n_{2}$, and such
that $n_{1}\cdot n_{2}=1$.
Lightcone coordinates are then defined as
\begin{eqnarray}
k^{\mu} & \equiv & k^{+}n_{1}^\mu+k^{-}n_{2}^\mu+k_{\perp}^{\mu},\label{eq:LightconeDef}\\
\nonumber \\
k\cdot l & \equiv & k^{+}l^{-}+k^{-}l^{+}+k_{\perp}\cdot l_{\perp}
  \equiv  k^{+}l^{-}+k^{-}l^{+}-(\boldsymbol{k}\cdot\boldsymbol{l}).
\end{eqnarray}
The metric tensor will be decomposed into its longitudinal and transverse
parts as
\begin{eqnarray}
g^{\mu\nu} & \equiv & n_{1}^{\mu}n_{2}^{\nu}+n_{1}^{\nu}n_{2}^{\mu}+g_{\perp}^{\mu\nu}.\label{eq:MetricDef}
\end{eqnarray}
}
\begin{eqnarray}
\left(\mathcal{A}_{\sigma}\right)^{c} & = & -ig\int d^{D}z_{0}\bar{u}\left(p_{q},z_{0}\right)\gamma^{\mu}t^{b}\varepsilon_{\mu}^{cb}\left(k,z_{0}\right)v\left(p_{\bar{q}},z_{0}\right), \label{eq:BasicSubamp}
\end{eqnarray}
where, with similar methods as in \cite{McLerran:1994vd,Balitsky:1995ub} one can derive the effective quark, antiquark and gluon lines as obtained for example in \cite{Boussarie:2014lxa,Boussarie:2016ogo}:
\begin{eqnarray}
\bar{u}\left(p_{q},\,z_{0}\right) & = & \frac{1}{2}\left(\frac{p_{q}^{+}}{2\pi}\right)^{\frac{d}{2}}\int d^{d}\boldsymbol{x}_{1}e^{ip_{q}^{+}\left(z_{0}^{-}-\frac{\left(\boldsymbol{x}_{1}-\boldsymbol{z}_{0}\right)^{2}}{2z_{0}^{+}}+i0\right)-i\boldsymbol{p}_{q}\cdot\boldsymbol{x}_{1}+i\frac{z_{0}^{+}}{2p_{q}^{+}}\left(m^{2}+i0\right)}\nonumber \\
 &  & \times\left(\frac{i}{z_{0}^{+}}\right)^{\frac{d}{2}}\bar{u}_{p_{q}}\gamma^{+}\left[U_{\boldsymbol{x}_{1}}\theta\left(-z_{0}^{+}\right)+\theta\left(z_{0}^{+}\right)\right]\left(\gamma^{-}-\frac{\Slash {x}_{1\perp}-\Slash {z}_{0\perp}}{z_{0}^{+}}+\frac{m}{p_{q}^{+}}\right),\label{eq:QuarkLine}
\end{eqnarray}

\begin{eqnarray}
v\left(p_{\bar{q}},\,z_{0}\right) & = & \frac{1}{2}\left(\frac{p_{\bar{q}}^{+}}{2\pi}\right)^{\frac{d}{2}}\int d^{d}\boldsymbol{x}_{2}e^{ip_{\bar{q}}^{+}\left(z_{0}^{-}-\frac{\left(\boldsymbol{x}_{2}-\boldsymbol{z}_{0}\right)^{2}}{2z_{0}^{+}}+i0\right)-i\boldsymbol{p}_{\bar{q}}\cdot\boldsymbol{x}_{2}+i\frac{z_{0}^{+}}{2p_{\bar{q}}^{+}}\left(m^{2}+i0\right)}\nonumber \\
 &  & \times\left(\frac{i}{z_{0}^{+}}\right)^{\frac{d}{2}}\left(\gamma^{-}-\frac{\Slash {x}_{2\perp}-\Slash {z}_{0\perp}}{z_{0}^{+}}-\frac{m}{p_{\bar{q}}^{+}}\right)\left[U_{\boldsymbol{x}_{2}}^{\dagger}\theta\left(-z_{0}^{+}\right)+\theta\left(z_{0}^{+}\right)\right]\gamma^{+}v_{p_{\bar{q}}},\label{eq:AntiquarkLine}
\end{eqnarray}

\begin{eqnarray}
\varepsilon_{\mu}^{ba}\left(k,\,z_{0}\right) & = & \left(\frac{k^{+}}{2\pi}\right)^{\frac{d}{2}}\int d^{d}\boldsymbol{x}_{0}e^{-ik^{+}\left(z_{0}^{-}-\frac{\left(\boldsymbol{x}_{0}-\boldsymbol{z}_{0}\right)^{2}}{2z_{0}^{+}}-i0\right)+i\boldsymbol{k}\cdot\boldsymbol{x}_{0}}\nonumber \\
 &  & \times\left(\frac{-i}{z_{0}^{+}}\right)^{\frac{d}{2}}\left(g_{\perp\mu\sigma}+\frac{x_{0\perp\sigma}-z_{0\perp\sigma}}{z_{0}^{+}}n_{2\mu}\right)\left[U_{\boldsymbol{x}_{0}}^{ba}\theta\left(z_{0}^{+}\right)+\delta^{ab}\theta\left(-z_{0}^{+}\right)\right]\varepsilon_{k\perp}^{\sigma}.\label{eq:GluonLine}
\end{eqnarray}
In the above, $m$ is the heavy quark mass and we have kept the transverse momentum of the incoming gluon $\boldsymbol{k}$. 
$U$ is the Wilson line which arises after the eikonal interaction with the target field $A^-$ 
\begin{eqnarray}
U_{\boldsymbol{x}} & \equiv & \mathcal{P}e^{ig\int_{-\infty}^\infty dx^{+}A^{-}(x)},\label{eq:LineDefPos}
\end{eqnarray}
We will use the conventions that lines with color indices
as superscrispts $U_{\boldsymbol{x}}^{ab}$ are in the adjoint representation
while lines without color indices are in the fundamental representation.

Integrating over the interaction point $z_0$, we get 
\beq
 \left(\mathcal{A}_{\sigma}\right)^{c} = \varepsilon_{k\perp}^{\sigma}\bar{u}_{p_{q}}\mathcal{M}_{\sigma}^{c}v_{p_{\bar{q}}},
\eeq
where
\begin{eqnarray}
\mathcal{M}_{\sigma}^{c} & = & -\frac{ig}{2}\left(2\pi\right)\delta\left(p_{q}^{+}+p_{\bar{q}}^{+}-k^{+}\right)\left(\frac{p_{q}^{+}p_{\bar{q}}^{+}k^{+}}{2\pi}\right)^{\frac{d}{2}}\nonumber \\
 &  & \times\int d^{d}\boldsymbol{x}_{1}d^{d}\boldsymbol{x}_{2}d^{d}\boldsymbol{x}_{0}e^{-i\boldsymbol{p}_{q}\cdot\boldsymbol{x}_{1}-i\boldsymbol{p}_{\bar{q}}\cdot\boldsymbol{x}_{2}+i\boldsymbol{k}\cdot\boldsymbol{x}_{0}}\delta\left(p_{q}^{+}\boldsymbol{x}_{1}+p_{\bar{q}}^{+}\boldsymbol{x}_{2}-k^{+}\boldsymbol{x}_{0}\right)\nonumber \\
 &  & \times\left[\frac{1}{z_{0}^{+}}\left(\gamma_{\perp\sigma}\Slash {x}_{12\perp}-2\frac{p_{\bar{q}}^{+}}{k^{+}}x_{12\perp\sigma}\right)-m\frac{k^{+}}{p_{q}^{+}p_{\bar{q}}^{+}}\gamma_{\perp\sigma}\right]\gamma^{+}\label{eq:Subamp1}\\
 &  & \times\left[\left(-i\right)^{\frac{d}{2}}\int_{-\infty}^{0}dz_{0}^{+}\left(-z_{0}^{+}\right)^{-\frac{d}{2}}\left(U_{\boldsymbol{x}_{1}}t^{c}U_{\boldsymbol{x}_{2}}^{\dagger}\right)e^{i\frac{k^{+}\left(m^{2}-i0\right)}{2p_{q}^{+}p_{\bar{q}}^{+}}z_{0}^{+}-i\frac{p_{q}^{+}p_{\bar{q}}^{+}\boldsymbol{x}_{12}^{2}}{2k^{+}}\frac{1+i0}{z_{0}^{+}}}\right.\nonumber \\
 &  & \left.+i^{\frac{d}{2}}\int_{0}^{+\infty}dz_{0}^{+}\left(z_{0}^{+}\right)^{-\frac{d}{2}}\left(t^{d}U_{\boldsymbol{x}_{0}}^{cd}\right)e^{i\frac{k^{+}\left(m^{2}+i0\right)}{2p_{q}^{+}p_{\bar{q}}^{+}}z_{0}^{+}-i\frac{p_{q}^{+}p_{\bar{q}}^{+}\boldsymbol{x}_{12}^{2}}{2k^{+}}\frac{1-i0}{z_{0}^{+}}}\right].\nonumber 
\end{eqnarray}
Using the integrals, for $Q,Z>0$:
\begin{eqnarray}
\int_{-\infty}^{0}dz^{+}\left(-z^{+}\right)^{-n}e^{i\left(Q-i0\right)z^{+}-i\frac{Z+i0}{z^{+}}} & = & 2\left(-i\right)^{1-n}\left(\sqrt{\frac{Z}{Q}}\right)^{1-n}K_{n-1}\left(2\sqrt{QZ}\right)\nonumber \\
\label{eq:BesselInts}\\
\int_{0}^{+\infty}dz^{+}\left(z^{+}\right)^{-n}e^{i\left(Q+i0\right)z^{+}-i\frac{Z-i0}{z^{+}}} & = & 2i^{1-n}\left(\sqrt{\frac{Z}{Q}}\right)^{1-n}K_{n-1}\left(2\sqrt{QZ}\right),\nonumber 
\end{eqnarray}
we get
\begin{eqnarray}
\mathcal{M}_{\sigma}^{c} & = & ig\delta\left(p_{q}^{+}+p_{\bar{q}}^{+}-k^{+}\right)\int d^{d}\boldsymbol{x}_{1}d^{d}\boldsymbol{x}_{2}d^{d}\boldsymbol{x}_{0}e^{-i\boldsymbol{p}_{q}\cdot\boldsymbol{x}_{1}-i\boldsymbol{p}_{\bar{q}}\cdot\boldsymbol{x}_{2}+i\boldsymbol{k}\cdot\boldsymbol{x}_{0}}\label{eq:Subamp2}\\
 &  & \times\left(2\pi\right)\delta\left(p_{q}^{+}\boldsymbol{x}_{1}+p_{\bar{q}}^{+}\boldsymbol{x}_{2}-k^{+}\boldsymbol{x}_{0}\right)\left(\frac{m\left(k^{+}\right)^{2}}{2\pi\left|\vec{x}_{12}\right|}\right)^{\frac{d}{2}}\left(U_{\boldsymbol{x}_{1}}t^{c}U_{\boldsymbol{x}_{2}}^{\dagger}-t^{d}U_{\boldsymbol{x}_{0}}^{cd}\right)\nonumber \\
 &  & \times\left[K_{\frac{d}{2}}\left(m\left|\boldsymbol{x}_{12}\right|\right)\left(\gamma_{\perp\sigma}\Slash {x}_{12\perp}-2\frac{p_{\bar{q}}^{+}}{k^{+}}x_{12\perp\sigma}\right)-i\left|\boldsymbol{x}_{12}\right|K_{\frac{d}{2}-1}\left(m\left|\boldsymbol{x}_{12}\right|\right)\gamma_{\perp\sigma}\right]\gamma^{+}.\nonumber 
\end{eqnarray}
In four dimensions and after a few simple changes of variables, this
can finally be rewritten as the known result:
\begin{eqnarray}
\mathcal{M}_{\sigma}^{c} & = & img\delta(K^{+}-k^{+})\int d^{2}\boldsymbol{b}d^{2}\boldsymbol{r}e^{-i(\boldsymbol{K}-\boldsymbol{k})\cdot\boldsymbol{b}+iq^{+}(\frac{\boldsymbol{k}}{K^{+}}-\frac{\boldsymbol{q}}{q^{+}})\cdot\boldsymbol{r}}\left(U_{\boldsymbol{b}+\frac{\boldsymbol{r}}{2}}t^{c}U_{\boldsymbol{b}-\frac{\boldsymbol{r}}{2}}^{\dagger}- t^{d}U_{\boldsymbol{b}}^{cd}\right)\label{eq:SubampFinal}\\
 &  & \times\left[\frac{\gamma_{\perp\sigma}\Slash {r}_{\perp}-(1-2\frac{q^{+}}{K^{+}})r_{\perp\sigma}}{\left|\boldsymbol{r}\right|}K_{1}\left(m\left|\boldsymbol{r}\right|\right)-iK_{0}\left(m\left|\boldsymbol{r}\right|\right)\gamma_{\perp\sigma}\right]\gamma^{+}.\nonumber 
\end{eqnarray}
Our next step is to convolute this amplitude with the transition probability to a quarkonium in the NRQCD approach where one expands the charmonium wavefunction as a series in powers of the
relative velocity of its constituents. As we have noted already, we shall be interested in $C$-even  quarkonia, $\eta$ and $\left(\chi_{J}\right)_{J=0,1,2}$. The corresponding projectors for the pseudoscalar $n^{2s+1}L_0$ and vector $n^{2s+1}L_1$ states are
\begin{eqnarray}
\Pi_{0} & \equiv & \frac{1}{\sqrt{8m^{3}}}\left(\frac{\Slash {K}}{2}-\Slash {q}-m\right)\gamma_{5}\left(\frac{\Slash {K}}{2}+\Slash {q}+m\right),\nonumber \\
\label{eq:Projectors}\\
\Pi_{1}^{\rho} & \equiv & \frac{1}{\sqrt{8m^{3}}}\left(\frac{\Slash {K}}{2}-\Slash {q}-m\right)\gamma^{\rho}\left(\frac{\Slash {K}}{2}+\Slash {q}+m\right).\nonumber 
\end{eqnarray}
Then the perturbative part of the $^{1}S_{0}$ wave and $^{3}P_{J}$
wave transitions read respectively:
\begin{eqnarray}
\mathcal{A}^{\sigma,c}\left(^{1}S_{0}\right) & \equiv & \left[\mathrm{Tr}\left(\Pi_{0}(\mathcal{M}^{\sigma})^{c}\right)\right]_{q=0}\label{eq:1S0def}
\nonumber \\
 & = & \sqrt{2m}g\delta(K^{+}-k^{+})K^{+}\int d^{2}\boldsymbol{r}d^{2}\boldsymbol{b}e^{-i(\boldsymbol{K}-\boldsymbol{k})\cdot\boldsymbol{b}}\left(U_{\boldsymbol{b}+\frac{\boldsymbol{r}}{2}}t^{c}U_{\boldsymbol{b}-\frac{\boldsymbol{r}}{2}}^{\dagger}-U_{\boldsymbol{b}}t^{c}U_{\boldsymbol{b}}^{\dagger}\right)\nonumber \\
 &  & \times\epsilon^{\sigma_{\perp}\mu_{\perp}+-}\frac{r_{\perp\mu}}{\left|\boldsymbol{r}\right|}K_{1}\left(m\left|\boldsymbol{r}\right|\right),\label{eq:1S0general}
\end{eqnarray}
and
\begin{eqnarray}
\mathcal{A}^{\sigma,c}\left(^{3}P_{J}\right) & \equiv & \varepsilon_{\left(J\right)}^{\rho\mu}\left[\frac{d}{dq^{\rho}}\mathrm{Tr}\left(\Pi_{1\mu}(\mathcal{M}^{\sigma})^{c}\right)\right]_{q=0}\label{eq:3PJdef}\\
\nonumber \\
 & = & ig\sqrt{2m}\varepsilon_{\left(J\right)\rho\mu}\delta(K^{+}-k^{+})\int d^{2}\boldsymbol{r}d^{2}\boldsymbol{b}e^{-i(\boldsymbol{K}-\boldsymbol{k})\cdot\boldsymbol{b}}\nonumber \\
 &  & \times\Biggl\{ r_{\perp\alpha}K_{0}\left(m\left|\boldsymbol{r}\right|\right)\left(K^{+}g_{\perp}^{\sigma\mu}-K_{\perp}^{\sigma}n_{2}^{\mu}\right) \label{eq:3PJgeneral}\\
 &  & \times\left[\left(g_{\perp}^{\alpha\rho}-\frac{k_{\perp}^{\alpha}}{K^{+}}n_{2}^{\rho}\right)\left(U_{\boldsymbol{b}+\frac{\boldsymbol{r}}{2}}t^{c}U_{\boldsymbol{b}-\frac{\boldsymbol{r}}{2}}^{\dagger}\right)-\left(g_{\perp}^{\alpha\rho}-\frac{K_{\perp}^{\alpha}}{K^{+}}n_{2}^{\rho}\right)\left(U_{\boldsymbol{b}}t^{c}U_{\boldsymbol{b}}^{\dagger}\right)\right]\nonumber \\
 &  & +\frac{r_{\perp\alpha}K_{1}\left(m\left|\boldsymbol{r}\right|\right)}{m\left|\boldsymbol{r}\right|}\left(U_{\boldsymbol{b}+\frac{\boldsymbol{r}}{2}}t^{c}U_{\boldsymbol{b}-\frac{\boldsymbol{r}}{2}}^{\dagger}-U_{\boldsymbol{b}}t^{c}U_{\boldsymbol{b}}^{\dagger}\right)\nonumber \\
 &  & \times\Bigl[K^{+}(g_{\perp}^{\rho\sigma}g_{\perp}^{\alpha\mu}-g_{\perp}^{\alpha\rho}g_{\perp}^{\sigma\mu})+(K^{\mu}g_{\perp}^{\alpha\sigma}+K_{\perp}^{\alpha}g_{\perp}^{\sigma\mu}-K_{\perp}^{\sigma}g_{\perp}^{\alpha\mu})n_{2}^{\rho}\nonumber \\
 &  & +(K_{\perp}^{\sigma}g_{\perp}^{\alpha\rho}-K_{\perp}^{\alpha}g_{\perp}^{\rho\sigma}-g_{\perp}^{\alpha\sigma}\frac{m^{2}}{K^{+}}n_{2}^{\rho})n_{2}^{\mu}\Bigr]\Biggr\} .\nonumber 
\end{eqnarray}

Next we use the fact that the modified Bessel functions
are peaked around 0 to expand the integrand  as 
\beq
\int d^2\boldsymbol{r} K_n(m|\boldsymbol{r})F(\boldsymbol{r}) \approx \int d^2\boldsymbol{r}\left[K_n(m|\boldsymbol{r}|)F(\boldsymbol{0}) + K_n(m|\boldsymbol{r}|) \partial^\mu_{\perp}\partial_\mu F(\boldsymbol{0})\right].
\eeq
and perform the $d^2\boldsymbol{r}$ integrals
\begin{eqnarray}
\int d^{2}\boldsymbol{r}K_{0}\left(m\left|\boldsymbol{r}\right|\right)r_{\perp\alpha}r_{\perp}^{\nu} & = & -\frac{4\pi}{m^{4}}g_{\perp\alpha}^{\nu}\nonumber \\
\label{eq:BesselInt}\\
\int d^{2}\boldsymbol{r}\frac{K_{1}\left(m\left|\boldsymbol{r}\right|\right)}{m\left|\boldsymbol{r}\right|}r_{\perp\alpha}r_{\perp}^{\nu} & = & -\frac{2\pi}{m^{4}}g_{\perp\alpha}^{\nu}.\nonumber 
\end{eqnarray}

Finally, we notice that the leading NRQCD contribution for $\eta$
mesons and for $\chi_{J}$ mesons are color singlet contributions for $^{1}S_{0}$ and for $^{3}P_{J}$ waves, respectively. Introducing the NRQCD long distance matrix elements (LDME's) and the color singlet projector 
\begin{equation}
\frac{\delta^{ij}}{N_{c}}\left\langle \mathcal{O}_{\eta}\left(^{1}S_{0}^1\right)\right\rangle ^{\frac{1}{2}}, \qquad \frac{\delta^{ij}}{N_{c}}\left\langle \mathcal{O}_{\chi_{J}}\left(^{3}P^1_{J}\right)\right\rangle ^{\frac{1}{2}},\label{eq:NRQCDLDME}
\end{equation}
($ij$ are color indices and the second superscript `1' denotes color singlet)
  we arrive at the  gluon-to-meson transition amplitudes: 
\begin{eqnarray}
\mathcal{A}^{\sigma,c}(\eta) & = & \frac{2g\pi}{m^{3}}\sqrt{2m}\delta(K^{+}-k^{+})\int d^{2}\boldsymbol{b}e^{-i(\boldsymbol{K}-\boldsymbol{k})\cdot\boldsymbol{b}}\frac{1}{N_{c}}\mathrm{Tr}\left[\left(\partial_{\nu}U_{\boldsymbol{b}}^{\dagger}\right)U_{\boldsymbol{b}}t^{c}\right]\nonumber \\
 &  & \times K^{+}\epsilon^{\sigma_{\perp}\nu_{\perp}+-}\left\langle \mathcal{O}_{\eta}\left(^{1}S_{0}^1\right)\right\rangle ^{\frac{1}{2}}\label{eq:EtaExp}
\end{eqnarray}
for $\eta$ transitions, and
\begin{eqnarray}
\mathcal{A}^{\sigma,c}(\chi_{J}) & = & \frac{2g\pi}{m^{3}}\sqrt{2m}\delta(K^{+}-k^{+})\int d^{2}\boldsymbol{b}e^{-i(\boldsymbol{K}-\boldsymbol{k})\cdot\boldsymbol{b}}\frac{1}{N_{c}}\mathrm{Tr}\left[\left(\partial_{\alpha}U_{\boldsymbol{b}}^{\dagger}\right)U_{\boldsymbol{b}}t^{c}\right]\nonumber \\
 &  & \times\frac{i}{m}\varepsilon_{\left(J\right)\rho\mu}\mathcal{P}^{\sigma\alpha\rho\mu}\left\langle \mathcal{O}_{\chi_{J}}\left(^{3}P_{J}^1\right)\right\rangle ^{\frac{1}{2}},    \label{eq:ChiJExp}
\end{eqnarray}
for $\chi_{J}$ transitions.  We used the transversity condition $K^{\rho}\varepsilon_{J\rho\mu}=0$
to obtain (\ref{eq:ChiJExp}).
The tensor structure in (\ref{eq:ChiJExp})
reads
\begin{eqnarray}
\mathcal{P}^{\sigma\alpha\mu\rho} & \equiv & K^{+}(g_{\perp}^{\sigma\mu}g_{\perp}^{\alpha\rho}+g_{\perp}^{\alpha\mu}g_{\perp}^{\rho\sigma})-(K_{\perp}^{\sigma}g_{\perp}^{\alpha\rho}+K_{\perp}^{\alpha}g_{\perp}^{\rho\sigma})n_{2}^{\mu}\nonumber \\
 &  & +\left[(K_{\perp}^{\alpha}-2k_{\perp}^{\alpha})g_{\perp}^{\sigma\mu}-K_{\perp}^{\sigma}g_{\perp}^{\alpha\mu}\right]n_{2}^{\rho}+\frac{2}{K^+}\left(k_{\perp}^{\alpha}K_{\perp}^{\sigma}-2m^{2}g_{\perp}^{\alpha\sigma}\right)n_{2}^{\mu}n_{2}^{\rho}.\label{eq:3PJtensor}
\end{eqnarray}
We shall use the following compact notation which summarizes the above results
\begin{eqnarray}
\mathcal{A}^{\sigma,c}({}^{2S+1}L_{J}) & \equiv & \frac{2g\pi}{m^{3}}\sqrt{2m}\delta(K^{+}-k^{+})\left\langle \mathcal{O}_{M}\left(^{2S+1}L_{J}^{1}\right)\right\rangle ^{\frac{1}{2}}\nonumber \\
 &  & \times\int d^{2}\boldsymbol{b}e^{-i(\boldsymbol{K}-\boldsymbol{k})\cdot\boldsymbol{b}}\frac{1}{N_{c}}\mathrm{Tr}\left[\left(\partial_{\alpha}U_{\boldsymbol{b}}^{\dagger}\right)U_{\boldsymbol{b}}t^{c}\right]\label{eq:ALJ}\\
 &  & \times\mathcal{P}^{\sigma\alpha}({}^{2S+1}L_{J}),\nonumber 
\end{eqnarray}
where the tensor structures read
\begin{eqnarray}
\mathcal{P}^{\sigma\alpha}({}^{3}S_{0}) & \equiv & K^{+}\epsilon^{\sigma_{\perp}\alpha_{\perp}+-}, \label{eq:3so}\\
\nonumber \\
\mathcal{P}^{\sigma\alpha}({}^{3}P_{J}) & \equiv & \frac{i}{m}\varepsilon_{\left(J\right)\rho\mu}\mathcal{P}^{\sigma\alpha\rho\mu}.\label{eq:3pj}
\end{eqnarray}
In Appendix A, we compute the single-inclusive cross section of $^3S_0$ and $^3P_J$ states in $pp$ (or $pA$) collisions, in order to check the compatibility our results with the single-inclusive $J/\psi$ production computed in~\cite{Kang:2013hta}.

\subsection{Double quarkonium production}

We are now ready to write down the generic   diffractive amplitude for the process $gg\rightarrow M_{1}\left(^{2S_{1}+1}L_{1J_{1}}\right)M_{2}\left(^{2S_{2}+1}L_{2J_{2}}\right)$. We basically square the $g\to M$ amplitude (\ref{eq:ALJ}) and project onto the color singlet state. First let us assume that the final state consists  of a $c\bar{c}$ quarkonium and a $b\bar{b}$ quarkonium. In this case the formula 
\beq
{\rm Tr} \left[U_{\boldsymbol{b}_{1}} t^a\left(\partial_{\alpha_{1}}U_{\boldsymbol{b}_{1}}^{\dagger}\right)  \right] {\rm Tr} \left[U_{\boldsymbol{b}_{2}} t^a  \left(  \partial_{\alpha_{2}}U_{\boldsymbol{b}_{2}}^{\dagger}\right) \right] = \frac{1}{2}{\rm Tr} \left[ \left(\partial_{\alpha_1}U_{\boldsymbol{b}_{1}}^{\dagger}\right)U_{\boldsymbol{b}_{1}}   \left(\partial_{\alpha_{2}}U_{\boldsymbol{b}_{2}}^{\dagger}\right)U_{\boldsymbol{b}_{2}} \right]  \label{direct}
\eeq
immediately gives the core structure of the WW GTMD (\ref{eq:GTMDdef}).  The full amplitude is
\begin{eqnarray}
 &  & \left(\mathcal{S}^{\sigma_{1}\sigma_{2}}\right)(M_{1},M_{2})\nonumber \\
 & = & \frac{4g^{2}\pi^{2}}{\sqrt{m_{1}^{5}m_{2}^{5}}}\frac{\delta(K_{1}^{+}-k_{1}^{+})\delta(K_{2}^{+}-k_{2}^{+})}{N_{c}^{2}\left(N_{c}^{2}-1\right)}\int d^{2}\boldsymbol{b}_{1}d^{2}\boldsymbol{b}_{2}e^{-i\left(\boldsymbol{K}_{1}-\boldsymbol{k}_{1}\right)\cdot\boldsymbol{b}_{1}-i\left(\boldsymbol{K}_{2}-\boldsymbol{k}_{2}\right)\cdot\boldsymbol{b}_{2}}\label{eq:AmpSingDif}\\
 &  & \times\left\langle P_{T}^{\prime}\left|\mathrm{Tr}\left[\left(\partial_{\alpha_{1}}U_{\boldsymbol{b}_{1}}^{\dagger}\right)U_{\boldsymbol{b}_{1}}\left(\partial_{\alpha_{2}}U_{\boldsymbol{b}_{2}}^{\dagger}\right)U_{\boldsymbol{b}_{2}}\right]\right|P_{T}\right\rangle \nonumber \\
 &  & \times\left\langle \mathcal{O}_{M_{1}}\left(^{2S_{1}+1}L_{1J_{1}}^{1}\right)\right\rangle ^{\frac{1}{2}}\left\langle \mathcal{O}_{M_{2}}\left(^{2S_{2}+1}L_{2J_{2}}^{1}\right)\right\rangle ^{\frac{1}{2}}\mathcal{P}^{\sigma_{1}\alpha_{1}}(M_{1})\mathcal{P}^{\sigma_{2}\alpha_{2}}(M_{2}).\nonumber  \label{dif}
\end{eqnarray}
A complication arises when the pair consists of  quarkonia with the same flavor. In this case, there exists an `exchange' diagram in which a quark from one gluon recombines with an antiquark from the other gluon and the remaining $q\bar{q}$ pair forms the second quarkonium.   In order for this process to occur,  the two quarks and antiquarks have to be all within the distance of order $1/m$. We thus cannot simply Taylor-expand the Wilson line separately in each gluon wavefunction as we have done above. Instead we must go back to the original expression and find the structure
\beq
&&{\rm Tr} \left[ U_{\boldsymbol{x}_{1}}t^{c}U_{\boldsymbol{x}_{2}}^{\dagger}U_{\boldsymbol{y}_{1}}t^{c}U_{\boldsymbol{y}_{2}}^{\dagger} \right]  \nonumber \\ 
&& =\frac{1}{2} {\rm Tr} \left[ U_{\boldsymbol{x}_{2}}^{\dagger}U_{\boldsymbol{y}_{1}}   \right] {\rm Tr}\left[ U_{\boldsymbol{y}_{2}}^{\dagger}  U_{\boldsymbol{x}_{1}} \right] -\frac{1}{2N_c} {\rm Tr} \left[ U_{\boldsymbol{x}_{1}}U_{\boldsymbol{x}_{2}}^{\dagger}U_{\boldsymbol{y}_{1}} U_{\boldsymbol{y}_{2}}^{\dagger}    \right] ,
\eeq
 which is not associated with the WW gluon distribution. While this may seem a problem, it is intuitively clear  that such a contribution is negligible in the limit of large quark mass.  The probability to find all the four quarks and antiquarks within a small area of order $1/m^2$ is power suppressed  compared to the `direct' contribution (see Fig~\ref{fig1}). However, this suppression  is difficult to see in the NRQCD framework where to leading order one just multiplies the partonic cross section by the constant LDMEs. We think this is an artifact of the NRQCD approach, and the exchange diagram will  be suppressed in a more complete treatment of the problem. Therefore, while (\ref{dif}) and similar results below are valid for the production of a different-flavor pair, strictly speaking, we think they can be  also used for a same-flavor pair up to small corrections.

Before leaving this section, we should comment on the  color-octet production mechanism. In NRQCD, quarkonia can be produced in a color-octet state \cite{Bodwin:1994jh}. It is well known that, in the collinear factorization framework, the color octet contribution actually dominates over the color-singlet contribution for $J/\psi$ production, and this is also the case for $\chi_c$ production at large transverse momentum~ \cite{Shao:2014fca}. In this paper, we instead focus on low transverse momentum quarkonia production $p_T \sim m$ where the color-singlet channel may actually dominate~\cite{Shao:2014fca}, although  the NRQCD factorization in this region needs further investigation to clarify this issue.\footnote{
From our derivation we find that the color-octet contribution is not 
sensitive to the WWW distribution that we are after. A cleaner approach approach to avoid this problem is to consider doubly diffractive events, $pp \to pM_1M_2p$ as in \cite{Bhattacharya:2018lgm} The cross section is then quartic in the proton Wigner distribution.} On the other hand, it has been found that $\eta_c$ production is always dominated by the color-singlet channel in the whole range of transverse momentum~\cite{Han:2014jya,{Butenschoen:2014dra}}.
Moreover, in the $k_T$-factorization approach, $\chi_c$ production is clearly dominated by the color-singlet channel  \cite{Hagler:2000dd,Baranov:2015yea}. Indeed,  in the $\chi_{J=1}$ production channel  which will be our main focus, the subprocess $gg^*\to \chi_{1}^{\rm{singlet}}$ is not forbidden by the Landau-Yang theorem because one of the gluons is off-shell in  the $k_T$ and hybrid factorizations, although it is forbidden in the collinear factorization.  As we are considering the forward production of quarkonia at low to moderate $p_T$, the use of $k_T$ or hybrid factorization is more appropriate. We thus concentrate on the color-singlet production mechanism in this paper, and leave the color-octet case for future work.

\section{Hybrid factorization with double scattering}

\subsection{Double gluon PDF} 

The 2 gluons-to-2 quarkonia production amplitude (\ref{eq:AmpSingDif}) is to be squared and convoluted with the double gluon distribution of the projectile proton with momentum $P_P$. In doing so, one has to be careful about the fact that the Lorentz indices ($\sigma_1\sigma_2$) of the gluons in the amplitude and the complex-conjugate amplitude can in general be different. This forces us to consider the most general double gluon PDF  \cite{Buffing:2017mqm}
\begin{align}
 & \mathcal{F}_{a_{1}a_{2}}\left(x_{1},x_{2},\boldsymbol{k}_{1},\boldsymbol{k}_{2},q\right) \label{eq:doublePDF} \\
 & =\frac{2}{x_{1}x_{2}P_{P}^{+}} \int d^{2}\boldsymbol{k}_{1}d^{2}\boldsymbol{k}_{2} \int\frac{dr^{-}}{2\pi}\frac{dz^{-}}{2\pi}\frac{d^{2}\boldsymbol{r}}{\left(2\pi\right)^{2}}\frac{d^{2}\boldsymbol{z}}{\left(2\pi\right)^{2}}e^{ix_{1}P_P^{+}r^{-}+ix_{2}P_P^{+}z^{-}-i\boldsymbol{k}_{1}\cdot\boldsymbol{r}-i\boldsymbol{k}_{2}\cdot\boldsymbol{z}}\nonumber \\
 & \times\int dy^{-}d^{2}\boldsymbol{y}\, e^{i \boldsymbol{q}\cdot\boldsymbol{y}}\left\langle P_P\left|\Pi_{a_{1}}^{ii^{\prime}}G^{+i^{\prime}}\left(-\frac{r}{2}\right)G^{+i}\left(\frac{r}{2}\right)\Pi_{a_{2}}^{jj^{\prime}}G^{+j^{\prime}}\left(y-\frac{z}{2}\right)G^{+j}\left(y+\frac{z}{2}\right)\right|P_P\right\rangle _{y^{+}=r^{+}=z^{+}=0},\nonumber 
\end{align}
where $a_{1,2}=\{g,\Delta g, \delta g\}$ and  $\Pi_{g}^{ii^{\prime}}\equiv\delta^{ii^{\prime}},$ $\Pi_{\Delta g}^{ii^{\prime}}\equiv i\epsilon^{ii^{\prime}},$
$\Pi_{\delta g}^{ii^{\prime}}\equiv\tau^{ii^{\prime},ll^{\prime}}\equiv\frac{1}{2}\left(\delta^{il}\delta^{i^{\prime}l^{\prime}}+\delta^{il^{\prime}}\delta^{i^{\prime}l}-\delta^{ii^{\prime}}\delta^{ll^{\prime}}\right)$
are respectively the unpolarized, longitudinally polarized and linearly
polarized projectors.
The momentum $\boldsymbol{q}$ is conjugate to the relative transverse coordinate of the two gluons, and its dependence cannot be completely eliminated~\cite{Buffing:2017mqm}.

\subsection{Polarization sum}
For unpolarized quarkonium production, we square the amplitude (\ref{eq:AmpSingDif}) and sum over quarkonium
polarizations: 
\begin{eqnarray}
\mathcal{P}^{\sigma\alpha}\left(^{3}S_{0}\right)\mathcal{P}^{\sigma^{\prime}\alpha^{\prime}}\left(^{3}S_{0}\right) & = & \left(K^{+}\right)^{2}\epsilon^{\sigma_{\perp}\alpha_{\perp}+-}\epsilon^{\sigma_{\perp}^{\prime}\alpha_{\perp}^{\prime}+-},\label{eq:3S0sum}\\
\nonumber \\
\mathcal{P}^{\sigma\alpha}\left(^{3}P_{J}\right)\mathcal{P}^{\sigma^{\prime}\alpha^{\prime}}\left(^{3}P_{J}\right) & = & \sum_{pol}\frac{\varepsilon_{\left(J\right)\rho\mu}\varepsilon_{\left(J\right)\rho^{\prime}\mu^{\prime}}}{m^{2}}\mathcal{P}^{\sigma\alpha\rho\mu}\mathcal{P}^{\sigma^{\prime}\alpha^{\prime}\rho^{\prime}\mu^{\prime}},\label{eq:3PJsum}
\end{eqnarray}
where the indices in the complex conjugate amplitude are denoted with a prime.  
This can be evaluated explicitly using the following result for $^{3}P_{J}$ polarization sums:
\begin{eqnarray}
\varepsilon_{\left(0\right)\rho\mu}\varepsilon_{\left(0\right)\rho^{\prime}\mu^{\prime}}^{\ast} & = & \frac{1}{3}\Pi_{\rho\mu}\Pi_{\rho^{\prime}\mu^{\prime}},\nonumber \\
\nonumber \\
\sum\varepsilon_{\left(1\right)\rho\mu}\varepsilon_{\left(1\right)\rho^{\prime}\mu^{\prime}}^{\ast} & = & \frac{1}{2}\left(\Pi_{\rho\rho^{\prime}}\Pi_{\mu\mu^{\prime}}-\Pi_{\rho\mu^{\prime}}\Pi_{\rho^{\prime}\mu}\right),\label{eq:3PJallsums}\\
\nonumber \\
\sum\varepsilon_{\left(2\right)\rho\mu}\varepsilon_{\left(2\right)\rho^{\prime}\mu^{\prime}}^{\ast} & = & \frac{1}{2}\left(\Pi_{\rho\rho^{\prime}}\Pi_{\mu\mu^{\prime}}+\Pi_{\rho\mu^{\prime}}\Pi_{\rho^{\prime}\mu}\right)-\frac{1}{3}\Pi_{\rho\mu}\Pi_{\rho^{\prime}\mu^{\prime}},\nonumber 
\end{eqnarray}
where 
\begin{equation}
\Pi_{\rho\mu}\equiv-g_{\mu\rho}+\frac{P_{\mu}P_{\rho}}{4m^{2}},\label{eq:3PJbasetensor}
\end{equation}
and the contractions
\begin{eqnarray}
\Pi_{\rho\mu}\Pi_{\rho^{\prime}\mu^{\prime}}\mathcal{K}^{\sigma\alpha\rho\mu}\mathcal{K}^{\sigma^{\prime}\alpha^{\prime}\rho^{\prime}\mu^{\prime}} & = & 9\left(K^{+}\right)^{2}g_{\perp}^{\alpha\sigma}g_{\perp}^{\alpha^{\prime}\sigma^{\prime}}, \nonumber
\\
\nonumber \\
\Pi_{\rho\rho^{\prime}}\Pi_{\mu\mu^{\prime}}\mathcal{P}^{\sigma\alpha\rho\mu}\mathcal{P}^{\sigma^{\prime}\alpha^{\prime}\rho^{\prime}\mu^{\prime}} & = & \left(K^{+}\right)^{2}\left[2g_{\perp}^{\alpha\alpha^{\prime}}g_{\perp}^{\sigma\sigma^{\prime}}+2g_{\perp}^{\alpha\sigma^{\prime}}g_{\perp}^{\alpha^{\prime}\sigma}+g_{\perp}^{\alpha^{\prime}\sigma^{\prime}}g_{\perp}^{\alpha\sigma}-g_{\perp}^{\sigma\sigma^{\prime}}\frac{\left(K_{\perp}^{\alpha}-k_{\perp}^{\alpha}\right)\left(K_{\perp}^{\alpha^{\prime}}-\ell_{\perp}^{\alpha^{\prime}}\right)}{m^{2}}\right],\nonumber \\
\nonumber \\
\Pi_{\rho\mu^{\prime}}\Pi_{\rho^{\prime}\mu}\mathcal{P}^{\sigma\alpha\rho\mu}\mathcal{P}^{\sigma^{\prime}\alpha^{\prime}\rho^{\prime}\mu^{\prime}} & = & \left(K^{+}\right)^{2}\left[2g_{\perp}^{\alpha\sigma^{\prime}}g_{\perp}^{\alpha^{\prime}\sigma}+2g_{\perp}^{\alpha\alpha^{\prime}}g_{\perp}^{\sigma\sigma^{\prime}}+g_{\perp}^{\alpha\sigma}g_{\perp}^{\alpha^{\prime}\sigma^{\prime}}\right]. \label{last} \\ \nonumber
\end{eqnarray}
The result is 
\begin{eqnarray}
\mathcal{P}^{\sigma\alpha}\left(^{3}P_{0}\right)\mathcal{P}^{\sigma^{\prime}\alpha^{\prime}}\left(^{3}P_{0}\right) & = & 3\frac{\left(K^{+}\right)^{2}}{m^{2}}g_{\perp}^{\alpha\sigma}g_{\perp}^{\alpha^{\prime}\sigma^{\prime}}\nonumber \\
\nonumber \\
\mathcal{P}^{\sigma\alpha}\left(^{3}P_{1}\right)\mathcal{P}^{\sigma^{\prime}\alpha^{\prime}}\left(^{3}P_{1}\right) & = & -\frac{\left(K^{+}\right)^{2}}{2m^{4}}\left(K_{\perp}^{\alpha}-k_{\perp}^{\alpha}\right)\left(K_{\perp}^{\alpha^{\prime}}-\ell_{\perp}^{\alpha^{\prime}}\right)g_{\perp}^{\sigma\sigma^{\prime}} \label{eq:AllSums}\\
\nonumber \\
\mathcal{P}^{\sigma\alpha}\left(^{3}P_{2}\right)\mathcal{P}^{\sigma^{\prime}\alpha^{\prime}}\left(^{3}P_{2}\right) & = & \frac{2\left(K^{+}\right)^{2}}{m^{2}}\left[g_{\perp}^{\alpha\alpha^{\prime}}g_{\perp}^{\sigma\sigma^{\prime}}-g_{\perp}^{\alpha\sigma}g_{\perp}^{\alpha^{\prime}\sigma^{\prime}}+g_{\perp}^{\alpha\sigma^{\prime}}g_{\perp}^{\alpha^{\prime}\sigma}-\frac{\left(K_{\perp}^{\alpha}-k_{\perp}^{\alpha}\right)\left(K_{\perp}^{\alpha^{\prime}}-\ell_{\perp}^{\alpha^{\prime}}\right)}{4m^{2}}g_{\perp}^{\sigma\sigma^{\prime}}\right].\nonumber 
\end{eqnarray}
Note that we wrote these quantities with distinct incoming gluon transverse
momenta in the amplitude $k_\perp$ and in the complex conjugate amplitude $\ell_\perp$. This is required for the proper use of the double PDF (\ref{eq:doublePDF}).

\subsection{Full cross section}

We now have all the machinery to finally compute the full differential cross section. This is straightforward but tedious, as we have to consider all possible Lorentz index structures, namely, ${\cal G}_{1,2,3,4}$  from (\ref{eq:GTMDdef-1}) and $a_{1,2}=\{g,\Delta g, \delta g\}$ in (\ref{eq:doublePDF}). The generic result is 
\begin{eqnarray}
 &  & \frac{d\sigma\left(M_{1},M_{2}\right)}{dY_{1}dY_{2}d^{2}\boldsymbol{\Delta}d^{2}\boldsymbol{K}}\nonumber \\
 & = & \frac{\alpha_{s}^{2}}{16m_{1}^{5}m_{2}^{5}N_{c}^{4}\left(N_{c}^{2}-1\right)^{2}}\left\langle \mathcal{O}_{M_{1}}\left(^{2S_{1}+1}L_{1J_{1}}^{1}\right)\right\rangle \left\langle \mathcal{O}_{M_{2}}\left(^{2S_{2}+1}L_{2J_{2}}^{1}\right)\right\rangle \nonumber \\
 &  & \times x_{1}x_{2}\int d^{2}\boldsymbol{q}\left[\delta^{ii^{\prime}}\delta^{jj^{\prime}}\mathcal{F}_{g,g}\left(x_{1},x_{2},\boldsymbol{q}\right)-i\delta^{ii^{\prime}}\epsilon^{jj^{\prime}}\mathcal{F}_{g,\Delta g}\left(x_{1},x_{2},\boldsymbol{q}\right)+2\delta^{ii^{\prime}}\tau^{jj^{\prime},nn^{\prime}}\mathcal{F}_{g,\delta g}^{nn^{\prime}}\left(x_{1},x_{2},\boldsymbol{q}\right)\right.\nonumber \\
 &  & -i\epsilon^{ii^{\prime}}\delta^{jj^{\prime}}\mathcal{F}_{\Delta g,g}\left(x_{1},x_{2},\boldsymbol{q}\right)-\epsilon^{ii^{\prime}}\epsilon^{jj^{\prime}}\mathcal{F}_{\Delta g,\Delta g}\left(x_{1},x_{2},\boldsymbol{q}\right)-2i\epsilon^{ii^{\prime}}\tau^{jj^{\prime},nn^{\prime}}\mathcal{F}_{\Delta g,\delta g}^{nn^{\prime}}\left(x_{1},x_{2},\boldsymbol{q}\right)\label{eq:CrossSectionGen}\\
 &  & \left.+2\delta^{jj^{\prime}}\tau^{ii^{\prime},mm^{\prime}}\mathcal{F}_{\delta g,g}^{mm^{\prime}}\left(x_{1},x_{2},\boldsymbol{q}\right)-2i\epsilon^{jj^{\prime}}\tau^{ii^{\prime},mm^{\prime}}\mathcal{F}_{\delta g,\Delta g}^{mm^{\prime}}\left(x_{1},x_{2},\boldsymbol{q}\right)+4\tau^{jj^{\prime},nn^{\prime}}\tau^{ii^{\prime},mm^{\prime}}\mathcal{F}_{\delta g,\delta g}^{mm^{\prime},nn^{\prime}}\left(x_{1},x_{2},\boldsymbol{q}\right)\right]\nonumber \\
 &  & \times\Pi_{1}^{ii^{\prime},kk^{\prime}}\left(M_{1}\right)\Pi_{2}^{jj^{\prime},\ell\ell^{\prime}}\left(M_{2}\right)x\mathcal{G}^{k\ell}\left(\boldsymbol{K}-\frac{\boldsymbol{q}}{2},\boldsymbol{\Delta}\right)x\mathcal{G}^{k^{\prime}\ell^{\prime}\ast}\left(\boldsymbol{K}+\frac{\boldsymbol{q}}{2},\boldsymbol{\Delta}\right),\nonumber 
\end{eqnarray}
where $x_{i}\equiv\frac{K_{i}^{+}}{P_{P}^{+}},$ $\boldsymbol{\Delta}\equiv\boldsymbol{K}_{1}+\boldsymbol{K}_{2},$ and from now on, 
$\boldsymbol{K}\equiv\frac{\boldsymbol{K}_{1}-\boldsymbol{K}_{2}}{2}$. 
$Y_{i}$ the rapidity of meson $M_{i}$, and the hard subparts $\Pi_{1,2}$ are given by

\begin{eqnarray}
\Pi_{1}^{ii^{\prime},kk^{\prime}}\left(\eta\right) & = & \delta^{ii^{\prime}}\delta^{kk^{\prime}}-\delta^{ik^{\prime}}\delta^{i^{\prime}k},\nonumber \\
\nonumber \\
\Pi_{1}^{ii^{\prime},kk^{\prime}}\left(\chi_{0}\right) & = & 3\frac{\delta^{ik}\delta^{i^{\prime}k^{\prime}}}{m_{1}^{2}}\label{eq:Pi1},\\
\nonumber \\
\Pi_{1}^{ii^{\prime},kk^{\prime}}\left(\chi_{1}\right) & = & \frac{\left(\boldsymbol{K}^{k}+\frac{\boldsymbol{\Delta}^{k}-\boldsymbol{q}^{k}}{2}\right)\left(\boldsymbol{K}^{k^{\prime}}+\frac{\boldsymbol{\Delta}^{k^{\prime}}+\boldsymbol{q}^{k^{\prime}}}{2}\right)}{2m_{1}^{4}}\delta^{ii^{\prime}},\nonumber \\
\nonumber \\
\Pi_{1}^{ii^{\prime},kk^{\prime}}\left(\chi_{2}\right) & = & \frac{2}{m_{1}^{2}}\left[\delta^{ii^{\prime}}\delta^{kk^{\prime}}-\delta^{ik}\delta^{i^{\prime}k^{\prime}}+\delta^{i^{\prime}k}\delta^{ik^{\prime}}+\frac{\left(\boldsymbol{K}^{k}+\frac{\boldsymbol{\Delta}^{k}-\boldsymbol{q}^{k}}{2}\right)\left(\boldsymbol{K}^{k^{\prime}}+\frac{\boldsymbol{\Delta}^{k^{\prime}}+\boldsymbol{q}^{k^{\prime}}}{2}\right)}{4m_{1}^{2}}\delta^{ii^{\prime}}\right],\nonumber 
\end{eqnarray}
for meson 1 and
\begin{eqnarray}
\Pi_{2}^{jj^{\prime},\ell \ell^{\prime}}\left(\eta\right) & = & \delta^{jj^{\prime}}\delta^{\ell \ell^{\prime}}-\delta^{j\ell^{\prime}}\delta^{j^{\prime}\ell},\nonumber \\
\nonumber \\
\Pi_{2}^{jj^{\prime},\ell \ell^{\prime}}\left(\chi_{0}\right) & = & 3\frac{\delta^{j\ell}\delta^{j^{\prime}\ell^{\prime}}}{m_{2}^{2}}\label{eq:Pi2},\\
\nonumber \\
\Pi_{2}^{jj^{\prime},\ell \ell^{\prime}}\left(\chi_{1}\right) & = & \frac{\left(\boldsymbol{K}^{\ell}-\frac{\boldsymbol{\Delta}^{\ell}-\boldsymbol{q}^{\ell}}{2}\right)\left(\boldsymbol{K}^{\ell^{\prime}}-\frac{\boldsymbol{\Delta}^{\ell^{\prime}}+\boldsymbol{q}^{\ell^{\prime}}}{2}\right)}{2m_{2}^{4}}\delta^{jj^{\prime}},\nonumber \\
\nonumber \\
\Pi_{2}^{jj^{\prime},\ell \ell^{\prime}}\left(\chi_{2}\right) & = & \frac{2}{m_{2}^{2}}\left[\delta^{jj^{\prime}}\delta^{\ell \ell^{\prime}}-\delta^{j\ell}\delta^{j^{\prime}\ell^{\prime}}+\delta^{j^{\prime}\ell}\delta^{j\ell^{\prime}}+\frac{\left(\boldsymbol{K}^{\ell}-\frac{\boldsymbol{\Delta}^{\ell}-\boldsymbol{q}^{\ell}}{2}\right)\left(\boldsymbol{K}^{\ell^{\prime}}-\frac{\boldsymbol{\Delta}^{\ell^{\prime}}+\boldsymbol{q}^{\ell^{\prime}}}{2}\right)}{4m_{2}^{2}}\delta^{jj^{\prime}}\right],\nonumber 
\end{eqnarray}
for meson 2. Interestingly, the intrinsic momentum $\boldsymbol{q}$ of the double PDF enters the argument of the WW GTMD. Physically, $\boldsymbol{q}$ is conjugate to the relative distance of the two quarkonia. More precisely, it is conjugate to $\frac{\boldsymbol{b}_1+\boldsymbol{b}'_1}{2} -\frac{\boldsymbol{b}_2+\boldsymbol{b}'_2}{2}$ where $\boldsymbol{b}_{1,2}$ is as in (\ref{eq:AmpSingDif}) and $\boldsymbol{b}'_{1,2}$ are the corresponding coordinates in the complex-conjugate amplitude.

\section{Explicit cross sections}

In the following, we will assume that $\left|\boldsymbol{K}\right|\gg\left|\boldsymbol{q}\right|$.
The dependence on $\boldsymbol{q}$ is then completely absorbed in
the projectile Double PDF. We will thus define the integrated double
PDFs 
\begin{equation}
\mathcal{F}_{a_{1},a_{2}}\left(x_{1},x_{2}\right)\equiv\int d^{2}\boldsymbol{q} \mathcal{F}_{a_{1},a_{2}}\left(x_{1},x_{2},\boldsymbol{q}\right).\label{eq:IntDPDF}
\end{equation}
At this point, we can already cancel some contributions. Indeed the symmetry and tracelessness properties of the $(a_1=g,a_2=\delta g)$ double PDF allows one to write it as:
\begin{equation}
\mathcal{F}_{g,\delta g}^{mn}\left(x_{1},x_{2},\boldsymbol{q}\right)=\left(\frac{\boldsymbol{q}^{m}\boldsymbol{q}^{n}}{\boldsymbol{q}^{2}}-\frac{\delta^{mn}}{2}\right)\frac{\boldsymbol{q}^{2}}{M_{P}^{2}}\mathcal{H}_{g,\delta g}\left(x_{1},x_{2},\boldsymbol{q}^{2}\right),\label{eq:Fgdg}
\end{equation}
where $M_P$ is the projectile's mass. It is then easy to show that the integral given in \ref{eq:IntDPDF} cancels for this integrated double PDF. Similarly, we can cancel the $(a_1=\delta g,a_2=g)$ integrated double PDF.

It will also be useful to write the $(a_1,a_2)=(\delta g,\delta g)$ integrated double PDF as
\begin{equation}
\mathcal{F}_{\delta g,\delta g}^{mm^{\prime},nn^{\prime}}\left(x_{1},x_{2}\right)\equiv\frac{1}{2}\left(\delta^{mn}\delta^{m^{\prime}n^{\prime}}+\delta^{mn^{\prime}}\delta^{m^{\prime}n}-\delta^{mm^{\prime}}\delta^{nn^{\prime}}\right)\mathcal{H}_{\delta g,\delta g}\left(x_{1},x_{2}\right).\label{eq:linlinPDF}
\end{equation}

\subsection{$\chi_1\chi_1$ cross section}

We find that the most concise formula is obtained for the double $\chi_{J=1}$ production.  It is given by 
\begin{eqnarray}
\frac{d\sigma\left(\chi_{f_11},\chi_{f_21}\right)}{dY_{1}dY_{2}d^{2}\boldsymbol{\Delta}d^{2}\boldsymbol{K}} & = & \frac{x_{1}x_{2}\mathcal{F}_{g,g}\left(x_{1},x_{2}\right)}{64m_{1}^{9}m_{2}^{9}N_{c}^{4}\left(N_{c}^{2}-1\right)^{2}}\alpha_{s}^{4}\left\langle \mathcal{O}_{\chi_{f_1 1}}\left(^{3}P_{1}^{1}\right)\right\rangle \left\langle \mathcal{O}_{\chi_{f_2 1}}\left(^{3}P_{1}^{1}\right)\right\rangle \nonumber \\
 &  & \times\left|\left(\boldsymbol{K}^{i}+\frac{\boldsymbol{\Delta}^{i}}{2}\right)\left(\boldsymbol{K}^{j}-\frac{\boldsymbol{\Delta}^{j}}{2}\right)x\mathcal{G}^{ij}\left(\boldsymbol{K},\boldsymbol{\Delta}\right)\right|^{2},\label{eq:Chi1Chi1}
\end{eqnarray}
with $f_{1,2}$ the meson flavors. To check that the right hand side has the correct dimensions, we note that ${\rm dim}\, {\cal G}=-2$,  ${\rm dim}\, {\cal F}=2$, and  ${\rm dim}\, \langle O_{\chi}\rangle = 5$ (${\rm dim}\, \langle O_{\eta}\rangle = 3$).   
Explicitly, the second line of (\ref{eq:Chi1Chi1}) reads

\begin{eqnarray}
 &  & \left|\left(\boldsymbol{K}^{i}+\frac{\boldsymbol{\Delta}^{i}}{2}\right)\left(\boldsymbol{K}^{j}-\frac{\boldsymbol{\Delta}^{j}}{2}\right)x\mathcal{G}^{ij}\left(\boldsymbol{K},\boldsymbol{\Delta}\right)\right|^{2}\nonumber \\
\nonumber \\
 & = & \boldsymbol{K}^{4}\left|\left(1-\frac{\boldsymbol{\Delta}^{2}}{4\boldsymbol{K}^{2}}\right)\mathcal{G}_{1}+\left(1-\frac{\left(\boldsymbol{K}\cdot\boldsymbol{\Delta}\right)^{2}}{2\boldsymbol{K}^{4}}+\frac{\boldsymbol{\Delta}^{2}}{4\boldsymbol{K}^{2}}\right)\frac{\boldsymbol{K}^{2}}{2M^{2}}\mathcal{G}_{2}\right.\label{eq:Chi1Chi1Tensor}\\
 &  & \left.-\left(\frac{\boldsymbol{\Delta}^{2}}{2M^{2}}-\frac{\left(\boldsymbol{K}\cdot\boldsymbol{\Delta}\right)^{2}}{\boldsymbol{K}^{2}M^{2}}+\frac{\boldsymbol{\Delta}^{4}}{8\boldsymbol{K}^{2}M^{2}}\right)\mathcal{G}_{3}-\left(\frac{\boldsymbol{\Delta}^{2}}{M^{2}}-\frac{\left(\boldsymbol{K}\cdot\boldsymbol{\Delta}\right)^{2}}{\boldsymbol{K}^{2}M^{2}}\right)\mathcal{G}_{4}\right|^{2}\nonumber \\
&\approx& \boldsymbol{K}^4\left({\cal G}_1+ \frac{\boldsymbol{K}^2}{2M^2}{\cal G}_2\right)^2,
\end{eqnarray}
where the last line is obtained by assuming $|\boldsymbol{K}|\gg |\boldsymbol{\Delta}|$. 
Remarkably, the cross section is directly proportional to the WW GTMD squared, without any convolution in momentum. 

\subsection{All averaged cross sections}

We will now present all the full cross sections for our $(M_1,M_2)$ process, for $ M_i\in \{\eta,\chi_0,\chi_1,\chi_2\}$. In order to get rid of the contributions with a longitudinally polarized double PDF, we will take the average w.r.t. the angle between $\mathbf{K}$ and $\mathbf{\Delta}$. This allows to keep only $\mathcal{F}_{g,g}(x_1,x_2)$ and $\mathcal{H}_{\delta g,\delta g}(x_1,x_2)$ as the non-perturbative distributions on the projectile side, since our goal is to focus rather on the non-perturbative effects on the target side.

\subsubsection{$\left(\chi_{1}\chi_{1}\right)$}
\vspace{-.5cm}
\begin{eqnarray}
&  &  \quad \quad \int_{0}^{2\pi}\frac{d\phi}{2\pi}\frac{d\sigma\left(\chi_{f_{1}1},\chi_{f_{2}1}\right)}{dY_{1}dY_{2}d^{2}\boldsymbol{K}d\boldsymbol{\Delta}^{2}}\nonumber \\
 &  & =\frac{\alpha_{s}^{4}x^{2}\boldsymbol{K}^{4}}{32m_{1}^{9}m_{2}^{9}N_{c}^{4}\left(N_{c}^{2}-1\right)^{2}}x_{1}x_{2}\mathcal{F}_{g,g}\left(x_{1},x_{2}\right)\left\langle \mathcal{O}_{\chi_{f_{1}1}}\left(^{3}P_{1}^{1}\right)\right\rangle \left\langle \mathcal{O}_{\chi_{f_{2}1}}\left(^{3}P_{1}^{1}\right)\right\rangle \label{eq:Chi1Chi1XS}\\
 &  & \times\left[\left(\mathcal{G}_{1}+\frac{\boldsymbol{K}^{2}}{2M^{2}}\mathcal{G}_{2}\right)^{2}-\frac{\boldsymbol{\Delta}^{2}}{2\boldsymbol{K}^{2}}\left(\mathcal{G}_{1}+\frac{\boldsymbol{K}^{2}}{2M^{2}}\mathcal{G}_{2}\right)\left(\mathcal{G}_{1}+2\frac{\boldsymbol{K}^{2}}{M^{2}}\mathcal{G}_{4}\right)\right].\nonumber 
\end{eqnarray}

\subsubsection{$\left(\chi_{1}\chi_{0}\right)$}
\vspace{-.5cm}
\begin{eqnarray}
 &  & \quad \quad \int_{0}^{2\pi}\frac{d\phi}{2\pi}\frac{d\sigma\left(\chi_{1},\chi_{0}\right)}{dY_{1}dY_{2}d^{2}\boldsymbol{K}d\boldsymbol{\Delta}^{2}}\nonumber \\
 &  & =\frac{3\alpha_{s}^{4}x^{2}\boldsymbol{K}^{2}}{32m_{1}^{9}m_{2}^{7}N_{c}^{4}\left(N_{c}^{2}-1\right)^{2}}x_{1}x_{2}\mathcal{F}_{g,g}\left(x_{1},x_{2}\right)\left\langle \mathcal{O}_{\chi_{1}}\left(^{3}P_{1}^{1}\right)\right\rangle \left\langle \mathcal{O}_{\chi_{0}}\left(^{3}P_{0}^{1}\right)\right\rangle \nonumber \label{eq:Chi1Chi0XS} \\
 &  & \times\left[\left(\mathcal{G}_{1}+\frac{\boldsymbol{K}^{2}}{2M^{2}}\mathcal{G}_{2}\right)^{2}+\frac{\boldsymbol{\Delta}^{2}}{4\boldsymbol{K}^{2}}\left(\mathcal{G}_{1}^{2}+\frac{\boldsymbol{K}^{4}}{4M^{4}}\left(\mathcal{G}_{2}^{2}-8\mathcal{G}_{2}\mathcal{G}_{4}+8\mathcal{G}_{4}^{2}\right)\right)\right]
\end{eqnarray}

\subsubsection{$\left(\chi_{1}\chi_{2}\right)$}
\vspace{-.5cm}
\begin{eqnarray}
 &  & \quad \quad \int_{0}^{2\pi}\frac{d\phi}{2\pi}\frac{d\sigma\left(\chi_{1},\chi_{2}\right)}{dY_{1}dY_{2}d^{2}\boldsymbol{K}d\boldsymbol{\Delta}^{2}}\nonumber \\
 &  & =\frac{\alpha_{s}^{4}\boldsymbol{K}^{2}x^{2}}{8m_{1}^{7}m_{2}^{7}N_{c}^{4}\left(N_{c}^{2}-1\right)^{2}}\left\langle \mathcal{O}_{\chi_{1}}\left(^{3}P_{1}^{1}\right)\right\rangle \left\langle \mathcal{O}_{\chi_{2}}\left(^{3}P_{2}^{1}\right)\right\rangle \label{eq:Chi1Chi2XS}\\
 &  & \times x_{1}x_{2}\mathcal{F}_{g,g}\left(x_{1},x_{2}\right)\left\{ \left(1+\frac{\boldsymbol{K}^{2}}{4m_{2}^{2}}\right)\left(\mathcal{G}_{1}+\frac{\boldsymbol{K}^{2}}{2M^{2}}\mathcal{G}_{2}\right)^{2}\right.\nonumber \\
 &  & \left.+\frac{\boldsymbol{\Delta}^{2}}{4\boldsymbol{K}^{2}}\left[\left(1+\frac{\boldsymbol{K}^{2}}{m_{2}^{2}}\right)\mathcal{G}_{1}^{2}+\frac{3\boldsymbol{K}^{4}}{4m_{2}^{2}M^{2}}\mathcal{G}_{1}\mathcal{G}_{2}+\frac{\boldsymbol{K}^{4}}{4M^{4}}\left(1+\frac{\boldsymbol{K}^{2}}{2m_{2}^{2}}\right)\mathcal{G}_{2}^{2}-2\frac{\boldsymbol{K}^{4}}{M^{4}}\left(\mathcal{G}_{2}-\mathcal{G}_{4}\right)\mathcal{G}_{4}\right]\right\} \nonumber 
\end{eqnarray}

\subsubsection{$\left(\chi_{1}\eta\right)$}
\vspace{-.5cm}
\begin{eqnarray}
 &  & \quad \quad \frac{1}{2\pi}\int_{0}^{2\pi}d\phi\frac{d\sigma\left(\chi_{1},\eta\right)}{dY_{1}dY_{2}d\boldsymbol{\Delta}^{2}d^{2}\boldsymbol{K}}\nonumber \\
 & = & \frac{\alpha_{s}^{4}x^{2}\boldsymbol{K}^{2}}{32m_{1}^{7}m_{2}^{5}N_{c}^{4}\left(N_{c}^{2}-1\right)^{2}}x_{1}x_{2}\mathcal{F}_{g,g}\left(x_{1},x_{2}\right)\left\langle \mathcal{O}_{\chi_{1}}\left(^{3}P_{1}^{1}\right)\right\rangle \left\langle \mathcal{O}_{\eta}\left(^{1}S_{0}^{1}\right)\right\rangle \label{eq:Chi1EtaXS}\\
 &  & \times\left[\left(\mathcal{G}_{1}+\frac{\boldsymbol{K}^{2}}{2M^{2}}\mathcal{G}_{2}\right)^{2}+\frac{\boldsymbol{\Delta}^{2}}{4\boldsymbol{K}^{2}}\left(\mathcal{G}_{1}^{2}+\frac{\boldsymbol{K}^{4}}{4M^{4}}\left(\mathcal{G}_{2}^{2}-8\mathcal{G}_{2}\mathcal{G}_{4}+8\mathcal{G}_{4}^{2}\right)\right)\right]\nonumber 
\end{eqnarray}

\subsubsection{$\left(\chi_{0}\chi_{0}\right)$}
\vspace{-.5cm}
\begin{eqnarray}
 &  & \quad \quad \int_{0}^{2\pi}\frac{d\phi}{2\pi}\frac{d\sigma\left(\chi_{f_{1}0},\chi_{f_{2}0}\right)}{dY_{1}dY_{2}d^{2}\boldsymbol{K}d\boldsymbol{\Delta}^{2}}\nonumber \\
 &  & =\frac{9\alpha_{s}^{4}x^{2}}{16 m_{1}^{7}m_{2}^{7}N_{c}^{4}\left(N_{c}^{2}-1\right)^{2}}\left\langle \mathcal{O}_{\chi_{f_{1}0}}\left(^{3}P_{0}^{1}\right)\right\rangle \left\langle \mathcal{O}_{\chi_{f_{2}0}}\left(^{3}P_{0}^{1}\right)\right\rangle \nonumber \\
 &  & \times\left\{ x_{1}x_{2}\mathcal{F}_{g,g}\left(x_{1},x_{2}\right)\left(\mathcal{G}_{1}^{2}+\frac{\boldsymbol{K}^{4}}{4M^{4}}\mathcal{G}_{2}^{2}+\frac{\boldsymbol{K}^{2}\boldsymbol{\Delta}^{2}}{2M^{4}}\mathcal{G}_{4}^{2}\right)\right.\label{eq:Chi0Chi0XS}\\
 &  & \left.+4x_{1}x_{2}\mathcal{H}_{\delta g,\delta g}\left(x_{1},x_{2}\right)\left[\mathcal{G}_{1}^{2}-2\frac{\boldsymbol{\Delta}^{2}}{M^{2}}\left(\mathcal{G}_{1}\mathcal{G}_{3}-\frac{\boldsymbol{K}^{2}}{4M^{2}}\mathcal{G}_{4}^{2}\right)\right]\right\} \nonumber 
\end{eqnarray}

\subsubsection{$\left(\chi_{0}\chi_{2}\right)$}
\vspace{-.5cm}
\begin{eqnarray}
 &  & \quad \quad \int_{0}^{2\pi}\frac{d\phi}{2\pi}\frac{d\sigma\left(\chi_{0},\chi_{2}\right)}{dY_{1}dY_{2}d^{2}\boldsymbol{K}d\boldsymbol{\Delta}^{2}}\nonumber \\
 &  & =\frac{3\alpha_{s}^{4}x^{2}}{16m_{1}^{7}m_{2}^{7}N_{c}^{4}\left(N_{c}^{2}-1\right)^{2}}x_{1}x_{2}\mathcal{F}_{g,g}\left(x_{1},x_{2}\right)\left\langle \mathcal{O}_{\chi_{0}}\left(^{3}P_{0}^{1}\right)\right\rangle \left\langle \mathcal{O}_{\chi_{2}}\left(^{3}P_{2}^{1}\right)\right\rangle \nonumber \\
 &  & \times\left[4\mathcal{G}_{1}^{2}+\frac{\boldsymbol{K}^{4}}{M^{4}}\mathcal{G}_{2}^{2}+\frac{\boldsymbol{K}^{2}}{2m_{2}^{2}}\left(\mathcal{G}_{1}+\frac{\boldsymbol{K}^{2}}{2M^{2}}\mathcal{G}_{2}\right)^{2}\right.\label{eq:Chi0Chi2XS}\\
 &  & \left.+\frac{\boldsymbol{\Delta}^{2}}{8M^{2}}\left(\frac{M^{2}}{m_{2}^{2}}\mathcal{G}_{1}^{2}+16\frac{\boldsymbol{K}^{2}}{M^{2}}\mathcal{G}_{4}^{2}+\frac{\boldsymbol{K}^{4}}{4m_{2}^{2}M^{2}}\left(\mathcal{G}_{2}^{2}+8\mathcal{G}_{2}\mathcal{G}_{4}+8\mathcal{G}_{4}^{2}\right)\right)\right]\nonumber 
\end{eqnarray}

\subsubsection{$\left(\chi_{0}\eta\right)$}
\vspace{-.5cm}
\begin{eqnarray}
 &  & \quad \quad \int_{0}^{2\pi}\frac{d\phi}{2\pi}\frac{d\sigma\left(\chi_{0},\eta\right)}{dY_{1}dY_{2}d^{2}\boldsymbol{K}d\boldsymbol{\Delta}^{2}}\nonumber \\
 &  & =\frac{3\alpha_{s}^{4}x^{2}}{16m_{1}^{7}m_{2}^{5}N_{c}^{4}\left(N_{c}^{2}-1\right)^{2}}\left\langle \mathcal{O}_{\chi_{0}}\left(^{3}P_{0}^{1}\right)\right\rangle \left\langle \mathcal{O}_{\eta}\left(^{1}S_{0}^{1}\right)\right\rangle \nonumber \\
 &  & \left\{ x_{1}x_{2}\mathcal{F}_{g,g}\left(x_{1},x_{2}\right)\left(\mathcal{G}_{1}^{2}+\frac{\boldsymbol{K}^{4}}{4M^{4}}\mathcal{G}_{2}^{2}+\frac{\boldsymbol{K}^{2}\boldsymbol{\Delta}^{2}}{2M^{4}}\mathcal{G}_{4}^{2}\right)\right.\label{eq: Chi0EtaXS}\\
 &  & \left.-4x_{1}x_{2}\mathcal{H}_{\delta g,\delta g}\left(x_{1},x_{2}\right)\left[\mathcal{G}_{1}^{2}-2\frac{\boldsymbol{\Delta}^{2}}{M^{2}}\left(\mathcal{G}_{1}\mathcal{G}_{3}-\frac{\boldsymbol{K}^{2}}{4M^{2}}\mathcal{G}_{4}^{2}\right)\right]\right\} \nonumber 
\end{eqnarray}

\subsubsection{$\left(\chi_{2}\chi_{2}\right)$}
\vspace{-.5cm}
\begin{eqnarray}
 &  & \quad \quad \int_{0}^{2\pi}\frac{d\phi}{2\pi}\frac{d\sigma\left(\chi_{f_{1}2},\chi_{f_{2}2}\right)}{dY_{1}dY_{2}d^{2}\boldsymbol{K}d\boldsymbol{\Delta}^{2}}\nonumber \\
 &  & =\frac{\alpha_{s}^{4}x^{2}}{8m_{1}^{7}m_{2}^{7}N_{c}^{4}\left(N_{c}^{2}-1\right)^{2}}x_{1}x_{2}\mathcal{F}_{g,g}\left(x_{1},x_{2}\right)\left\langle \mathcal{O}_{\chi_{f_{1}2}}\left(^{3}P_{2}^{1}\right)\right\rangle \left\langle \mathcal{O}_{\chi_{f_{2}2}}\left(^{3}P_{2}^{1}\right)\right\rangle \nonumber \\
 &  & \times\left\{ 8\left(\mathcal{G}_{1}^{2}+\frac{\boldsymbol{K}^{4}}{4M^{4}}\mathcal{G}_{2}^{2}\right)+\left(\frac{\boldsymbol{K}^{2}}{m_{1}^{2}}+\frac{\boldsymbol{K}^{2}}{m_{2}^{2}}+\frac{\boldsymbol{K}^{4}}{4m_{1}^{2}m_{2}^{2}}\right)\left(\mathcal{G}_{1}+\frac{\boldsymbol{K}^{2}}{2M^{2}}\mathcal{G}_{2}\right)^{2}\right.\nonumber \\
 &  & +\frac{\boldsymbol{\Delta}^{2}}{32m_{1}^{2}m_{2}^{2}}\left[8\mathcal{G}_{1}^{2}\left(\boldsymbol{K}^{2}+m_{1}^{2}+m_{2}^{2}\right)+6\boldsymbol{K}^{2}\frac{\boldsymbol{K}^{2}}{M^{2}}\mathcal{G}_{1}\mathcal{G}_{2}+\frac{\boldsymbol{K}^{4}}{M^{4}}\left(\boldsymbol{K}^{2}+2m_{1}^{2}+2m_{2}^{2}\right)\mathcal{G}_{2}^{2}\right]\label{eq:Chi2Chi2XS}\\
 &  & \left.+\frac{\boldsymbol{\Delta}^{2}}{2m_{1}^{2}m_{2}^{2}}\frac{\boldsymbol{K}^{2}}{M^{2}}\left[\left(m_{1}^{2}-m_{2}^{2}\right)\frac{\boldsymbol{K}^{2}}{M^{2}}\mathcal{G}_{2}\mathcal{G}_{4}+\left(\frac{\boldsymbol{K}^{2}m_{1}^{2}}{M^{2}}+\frac{\boldsymbol{K}^{2}m_{2}^{2}}{M^{2}}+8\frac{m_{1}^{2}m_{2}^{2}}{M^{2}}\right)\mathcal{G}_{4}^{2}\right]\right\} \nonumber 
\end{eqnarray}

\subsubsection{$\left(\chi_{2}\eta\right)$}
\vspace{-.5cm}
\begin{eqnarray}
 &  & \quad \quad \int_{0}^{2\pi}\frac{d\phi}{2\pi}\frac{d\sigma\left(\chi_{2},\eta\right)}{dY_{1}dY_{2}d^{2}\boldsymbol{K}d\boldsymbol{\Delta}^{2}}\nonumber \\
 &  & =\frac{\alpha_{s}^{4}x^{2}}{16m_{1}^{7}m_{2}^{5}N_{c}^{4}\left(N_{c}^{2}-1\right)^{2}}\left\langle \mathcal{O}_{\chi_{2}}\left(^{3}P_{2}^{1}\right)\right\rangle \left\langle \mathcal{O}_{\eta}\left(^{1}S_{0}^{1}\right)\right\rangle \label{eq:Chi2EtaXS}\\
 &  & \times x_{1}x_{2}\mathcal{F}_{g,g}\left(x_{1},x_{2}\right)\left[4\left(\mathcal{G}_{1}^{2}+\frac{\boldsymbol{K}^{4}}{4M^{4}}\mathcal{G}_{2}^{2}\right)+\frac{\boldsymbol{K}^{2}}{2m_{1}^{2}}\left(\mathcal{G}_{1}+\frac{\boldsymbol{K}^{2}}{2M^{2}}\mathcal{G}_{2}\right)^{2}\right.\nonumber \\
 &  & \left.+\frac{\boldsymbol{\Delta}^{2}}{8m_{1}^{2}}\left(\mathcal{G}_{1}^{2}+\frac{\boldsymbol{K}^{4}}{4M^{4}}\left(\mathcal{G}_{2}^{2}-8\mathcal{G}_{2}\mathcal{G}_{4}+8\mathcal{G}_{4}^{2}\right)+16\frac{\boldsymbol{K}^{2}}{M^{2}}\frac{m_{1}^{2}}{M^{2}}\mathcal{G}_{4}^{2}\right)\right]\nonumber 
\end{eqnarray}

\subsubsection{$\left(\eta\eta\right)$}
\vspace{-.5cm}
\begin{eqnarray}
 &  & \quad \quad \int_{0}^{2\pi}\frac{d\phi}{2\pi}\frac{d\sigma\left(\eta_{f_{1}},\eta_{f_{2}}\right)}{dY_{1}dY_{2}d^{2}\boldsymbol{K}d\boldsymbol{\Delta}^{2}}\nonumber \\
 &  & =\frac{\alpha_{s}^{4}x^{2}}{16m_{1}^{5}m_{2}^{5}N_{c}^{4}\left(N_{c}^{2}-1\right)^{2}}\left\langle \mathcal{O}_{\eta_{f_{1}}}\left(^{1}S_{0}^{1}\right)\right\rangle \left\langle \mathcal{O}_{\eta_{f_{2}}}\left(^{1}S_{0}^{1}\right)\right\rangle \nonumber \\
 &  & \times\left\{ x_{1}x_{2}\mathcal{F}_{g,g}\left(x_{1},x_{2}\right)\left(\mathcal{G}_{1}^{2}+\frac{\boldsymbol{K}^{4}}{4M^{4}}\mathcal{G}_{2}^{2}+\frac{\boldsymbol{K}^{2}\boldsymbol{\Delta}^{2}}{2M^{4}}\mathcal{G}_{4}^{2}\right)\right.\label{eq:EtaEtaXS}\\
 &  & \left.+4x_{1}x_{2}\mathcal{H}_{\delta g,\delta g}\left(x_{1},x_{2}\right)\left[\mathcal{G}_{1}^{2}-2\frac{\boldsymbol{\Delta}^{2}}{M^{2}}\left(\mathcal{G}_{1}\mathcal{G}_{3}-\frac{\boldsymbol{K}^{2}}{4M^{2}}\mathcal{G}_{4}^{2}\right)\right]\right\} .\nonumber 
\end{eqnarray}

\section{Conclusions}

In summary, in this paper we derive the diffractive double quarkonia production in $pp$ and $pA$ collisions to probe the WW gluon GTMDs. In particular, we applied the double parton scattering mechanism from the projectile, where the two gluons scatter off the nucleon/nucleus target diffractively to produce the final state two quarkonium states. The amplitudes are found to be sensitive to the WW gluon GTMDs.

More importantly, we found that the differential cross sections can be much simplified if we integrate out the transverse momenta of the gluons from the projectile. The explicit expressions show that the cross sections can be written as squared of the WW gluon GTMDs from the target.

Our result is a first example of direct access to the WW gluon GTMDs in hard diffractive processes. Experimentally, this process may be a challenge to measure. We hope that our derivations will stimulate further theoretical developments to explore the physics of GTMD and possibility to measure them in experiments.

\acknowledgments

This work is supported by the National Science Center, Poland, grant No. 2015/17/B/ST2/01838. It is also partially supported by the Natural Science Foundation of China (NSFC) under Grant Nos.~11575070 and by the U.S. Department of Energy, Office of Science, Office of Nuclear Physics, under contract number DE-AC02-05CH11231.

\appendix

\section{Gluon to $^1S_0$ and $^3P_J$ waves transition in the hybrid formalism}

For completeness, here we  write down the cross section for the single-inclusive quarkonium production  $gp\to MX$ for $^1S_0$ and $^3P_J$ waves. Squaring (\ref{eq:3pj}) and using the projectors (\ref{eq:3S0sum})--(\ref{last}), we find
\begin{eqnarray*}
\frac{d\sigma_{g}(M)}{dK^{+}d^{2}\boldsymbol{K}} & = & \frac{\alpha_{s}}{2m^{5}N_{c}\left(N_{c}^{2}-1\right)}\delta(K^{+}-k^{+})\int\frac{d^{2}\boldsymbol{b}d^{2}\boldsymbol{b}^{\prime}}{\left(2\pi\right)^{2}}e^{i\boldsymbol{K}\cdot(\boldsymbol{b}^{\prime}-\boldsymbol{b})}\\
 &  & \times\left\langle \frac{1}{N_{c}}\mathrm{Tr}\left[U_{\boldsymbol{b}}^{\dagger}(\partial_{i}U_{\boldsymbol{b}})U_{\boldsymbol{b}^{\prime}}^{\dagger}(\partial_{j}U_{\boldsymbol{b}^{\prime}})\right]\right\rangle \varphi^{ij}(^{2S+1}L_{J})\left\langle {\cal O}_M(^{2S+1}L_J^1)\right\rangle^{\frac{1}{2}} ,
\end{eqnarray*}

where

\begin{eqnarray}
\varphi^{ij}(^{1}S_{0}) & = & \delta^{ij},\nonumber \\
\nonumber \\
\varphi^{ij}(^{3}P_{0}) & = & \frac{3\delta^{ij}}{m^{2}},\label{eq:AllSumsFinal-1}\\
\nonumber \\
\varphi^{ij}(^{3}P_{1}) & = & \frac{K^{i}K^{j}}{m^{4}},\nonumber \\
\nonumber \\
\varphi^{ij}(^{3}P_{2}) & = & \frac{4}{m^{2}}\left(\delta^{ij}+\frac{K^{i}K^{j}}{4m^{2}}\right),\nonumber 
\end{eqnarray}
and the brackets denote the forward matrix element $\frac{\langle P|...|P\rangle}{\langle P|P\rangle}$, or equivalently the CGC averaging. \\
Taking the case where $M=J/\psi$ and taking the small-dipole limit in the collinear limit of results described in \cite{Kang:2013hta} shows full compatibility of the present results with previous calculations.

\end{document}